\documentclass[aps,pra,superscriptaddress,twocolumn]{revtex4}
\usepackage{amsmath,epsfig,mathrsfs,graphicx,amssymb,color}
\usepackage[T1]{fontenc}
\usepackage{t1enc}
\usepackage{hyperref}
\usepackage{array}
\usepackage{booktabs}
\usepackage{yquant}
\useyquantlanguage{qasm}
\usepackage{physics}

\def\mb{\begin{pmatrix}}
\def\me{\end{pmatrix}}
\def\be#1\ee{\begin{equation}#1\end{equation}}
\newcommand{\ba}{\begin{eqnarray} }
\newcommand{\ea}{\end{eqnarray} }

\begin{document}

\title{Precise certification of a qubit space}

\affiliation{Faculty of Physics, University of Warsaw, ul. Pasteura 5, PL02-093 Warsaw, Poland}
\affiliation{Systems Research Institute, Polish Academy of Sciences, 6 Newelska Street, PL01-447 Warsaw, Poland}
\affiliation{Center for Theoretical Physics, Polish Academy of Sciences, Al. Lotnik{\'o}w 32/46, PL02-668 Warsaw, Poland}
\affiliation{CRISP - Centre de Recerca Independent de sa Pobla, 07420 sa Pobla, Balearic Islands, Spain}
                
\author{Tomasz Bia{\l}ecki$^{1}$}
\author{Tomasz Rybotycki$^{2,3}$}
\author{Josep Batle$^{4}$}
\author{Jakub Tworzyd{\l}o$^{1}$}
\author{Adam Bednorz$^{1}$}

\email{Adam.Bednorz@fuw.edu.pl}

%\date{\today}

\begin{abstract}
We demonstrate an implementation of the precise test of dimension on the qubit, using the public IBM quantum computer,
using the determinant dimension witness.
The accuracy is below $10^{-3}$ comparing to maximal possible value of the witness in higher dimension.
The test involving minimal independent sets of preparation and measurement operations (gates) is applied both for specific configurations and
parametric ones. The test is be robust against nonidealities such as incoherent leakage and erroneous gate execution.
Two of the IBM devices failed the test by more than $5$ standard deviations, which has no simple explanation.

\end{abstract}

\maketitle

\section{Introduction}

Physics is an exact science, which is confirmed by precise measurements of fundamental constants and establishing definition of SI units
by precise quantum experiments \cite{gravity,grav2,grav3,fine,particle,photon}. Precision is also required from every computer, also quantum. Unfortunately, current quantum technologies suffer
from inevitable sources of errors, both just from mechanical limitations and inseparable physical environment. Of course, there are methods
to mitigate and correct the errors. Such approach relies, however, on assumptions about the controllable space of possible actions.

The basic building block of a quantum computer is a qubit, a generic two-level system. Since the goal is to manipulate accurately many qubits,
it is necessary to 
ascertain whether or not 
the qubit space is reliable, i.e. not combined with a larger space. The most promising implementations of qubits keep
them detuned from environment and other states, except for small incoherent disturbance. On the other hand, 
the  potential contribution of external states can lead to systematic errors, hard to correct.
Operations on qubits, gates, realized by microwave pulses, suffer from distortions due to nonlinearities of waveform generators \cite{distor},
so a simple deviation of the probability distribution from the theoretical prediction is not yet a proof of extra space \cite{praca}.
Therefore,  to increase the quality of classical and quantum computation and communication,
these systems need precise certification, robust against imperfections of physical implementations.

\begin{figure}
\includegraphics[scale=.3]{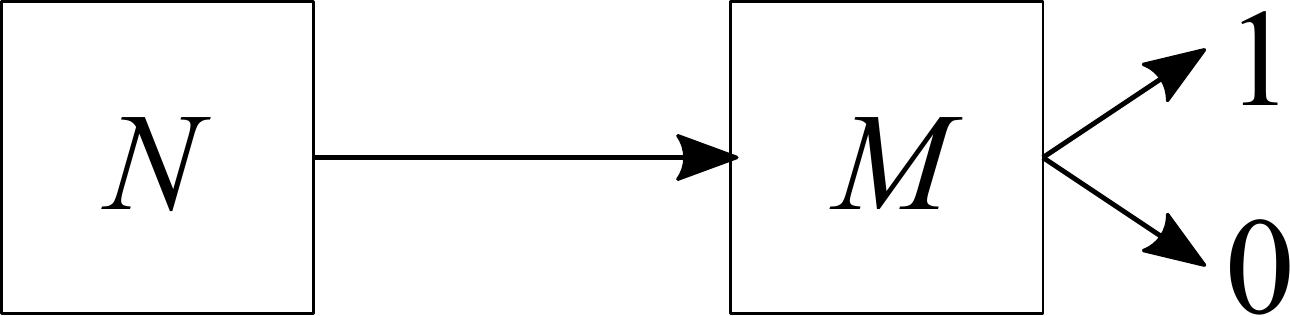}
\caption{Preparation and measurement scenario; the state is prepared as $N$ and measured by $M$ to give an outcome of either $1$ or $0$.}
\label{pms}
\end{figure}

The dimension of the quantum space can be checked by a dimension witness \cite{gallego,hendr,ahr,ahr2,dim1,leak}. 
The construction of the witness is based on the two-stage protocol, the initial preparation and subsequent final measurement, 
which are chosen from independent sets. The preparation must be completed before the start of the measurement.
 A precise witness must be based on equality, i.e.
a quantity, which is exactly zero up to a certain dimension, and nonzero otherwise. 
Such a good witness test is the linear independence of the  specific dichotomic outcome probability $p(M|N)$ for the preparation $N$ and measurement $M$, 
see Fig. \ref{pms}, tested by a suitable determinant \cite{dim, chen,bb22}. It has been been already performed on optical states \cite{opt}.
It belongs to a family to equality-based tests, like the Sorkin equality \cite{sorkin} in the three-slit experiment \cite{tslit,btest1,btest2} 
 testing Born's rule \cite{born}, benchmarking our trust in fundamental quantum models and their actual realizations.
 
 In this paper, we apply the test to several IBM quantum device. While some results agree with the $2-$level model, taking a large statistics revealed
 signature of the failure by more than 5 standard deviations. Of course it does not immediately mean a larger space but the problem needs urgent 
further investigation to determine the cause, which may be also another assumption of the test (e.g. lack of independence of the operations).

\section{Theory}
We apply a test of the qubit space $d=2$ with the witness constructed for $p(M|N)=\mathrm{tr}MN$, $N=N^\dag\geq 0$, 
$\mathrm{tr}N=1$ and measurement 
$1\geq M=M^\dag\geq 0$.
Taking $5$ preparations $N_j$, $j=1..5$ and $4$ measurements $M_k$, $k=1..4$.
Then the determinant $W=\det p$, for the $5\times 5$ matrix $p$ with entries $p_{kj}=p(M_j|N_k)$ and $p_{5j}=1$,
must be equal to zero if all $N_j$ and $M_k$
are represented in the same two-level space. In addition, it remains zero also if all preparations and measurements 
contain some constant incoherent leakage term, i.e.
$N'_j=N_j+N_e$ and $M'_k=M_k+M_e$, with $N_e$ and $M_e$ independent of $j$ and $k$ and commuting with $N_j$
and $M_k$. In this way, the common leakage to higher states does not affect the test \cite{leak}. For $d=2$ we have $W=0$, but $d=3$ gives
maximally $27\sqrt{2}/64\simeq 0.6$ in the real space and $\simeq 0.632$ in the complex space \cite{bb22}. For $d=4$ the maximum (real and complex) 
is $2^{12}/3^7\simeq 1.87$. Even higher dimensions are saturated by the classical maximum $3$.

The IBM Quantum Experience cloud computing offers several devices, collections of qubits, which can be manipulated
by a user-defined set of gates (operations) -- either single qubit or two-qubit ones, also paramteric.
One can put barriers (controlling the order of operations) or additional resets (nonunitary transition to the ground state). 
The qubits are physical transmons \cite{transmon}, the artificial quantum states existing due to interplay of superconductivity (Josephson effect) and capacitance. 
Due to anharmonicity one can limit the working space to two states.
The decoherence time  (mostly  environmental) is long enough to perform a sequence of quantum operations and read out reliable results. 

The ground state $|0\rangle$ can be additionally assured by a reset operation. 
Gates are implemented by time-scheduled microwave pulses prepared by waveform generators  and mixers 
(time $30-70$ns with sampling at $0.222$ns), tuned to the drive frequency  (energy difference between qubit levels) \cite{qis} (about $4-5$Ghz).
The rotation $Z$ is not a real pulse, but an realized by  an instantaneous virtual gate $\mathrm{VZ}(\theta)$, which adds a rotation 
between in- and out-of-phase components of the next gates \cite{zgates}.
The readout is performed  another  long microwave pulse of frequency  (different from the drive) to the resonator
and measuring the populated photons \cite{qis,read}.

In the following, we assume the two-level description of the qubits, expecting $W=0$ up to statistical error. Larger deviation
would be an evidence that this description is inaccurate.
The states and operators will be can be described either
in a two-dimensional Hilbert space with basis $|0\rangle$, $|1\rangle$ or in the Bloch sphere with $V=(v_0+\boldsymbol v\cdot\boldsymbol \sigma)/2$,
with the 3-component Bloch vector $\boldsymbol v$ and standard Pauli matrices 
\begin{align}
&\sigma_1=\begin{pmatrix}
0&1\\
1&0\end{pmatrix},\;\sigma_2=\begin{pmatrix}
0&-i\\
i&0\end{pmatrix},\;\sigma_3=\begin{pmatrix}
1&0\\
0&-1\end{pmatrix}.
\end{align}
Then the initial state $|0\rangle\langle 0|$ corresponds to the vector $(0,0,1)$ while $n_0=1$ and $|\boldsymbol n|\leq 1$ and 
$2-|\boldsymbol m|\geq m_0\geq |\boldsymbol m|$.
A microwave pulse tuned to the interlevel drive frequency corresponds to parametric gates, $\pi/2$ rotations,
\begin{align}
&S_\gamma=Z^\dag_\gamma SZ_\gamma ,\:
Z_\theta=
\begin{pmatrix}
e^{-i\gamma/2}&0\\
0&e^{i\gamma/2}\end{pmatrix},\nonumber\\
&
S=RX(\pi/2)=\sqrt{X}=\frac{1}{\sqrt{2}}\begin{pmatrix}
1&-i\\
-i&1\end{pmatrix}\label{smat},
\end{align}
in the basis $|0\rangle$, $|1\rangle$ while
\begin{equation}
S=\begin{pmatrix}
1&0&0\\
0&0&-1\\
0&1&0
\end{pmatrix},\;Z_\gamma=\begin{pmatrix}
\cos\gamma&-\sin\gamma&0\\
\sin\gamma&\cos\gamma&0\\
0&0&1\end{pmatrix},
\end{equation}
on the Bloch vector, i.e. $S_\gamma VS^\dag_\gamma$. 

\begin{figure}
	\centering
\begin{tikzpicture}[scale=1.3]
		\begin{yquant*}
			% q[1];
			init {$\ket 0$} q[0];
			% cbit c[1];
			box {$S_\alpha$} q[0];
			box {$S_{\beta}$} q[0];
			barrier q[0]; 
			box {$S_{\phi}$} q[0];
			box {$S_{\theta}$} q[0];
			measure q[0];
		\end{yquant*}
	\end{tikzpicture}
\caption{The quantum circuit for the dimension test. The initial state $|0\rangle$ and four gates $S_\gamma$, split into preparation and measurement
stages, are followed by the final dichotomic measurement}
\label{gates}
\end{figure}
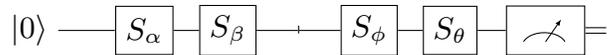

\begin{figure}
\includegraphics[scale=.8]{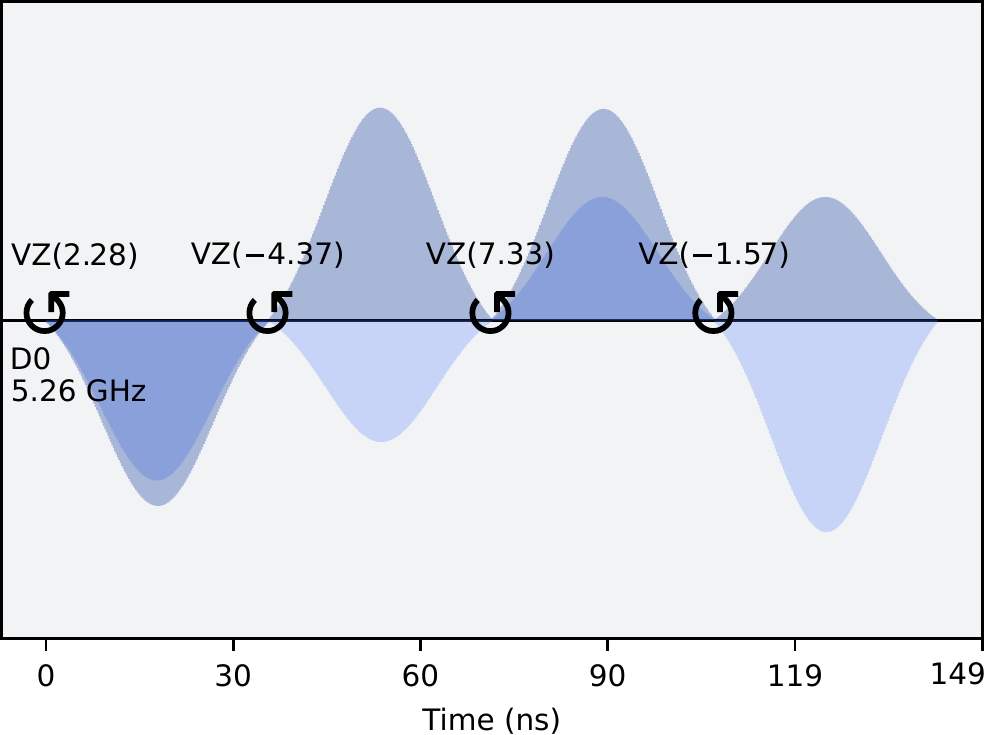}
\caption{The actual waveform of the pulse on IBM quantum computer (nairobi), with four subsequent gates $S_\gamma$, with $\gamma=\alpha,\beta,\phi,\theta$,
consecutively.
The discretization unit time is $dt=0.222$ns. 
Driving (level gap) frequency is denoted by $D0$. The light/dark shading corresponds to in-phase/out-of-phase amplitude component, respectively. 
The element $\mathrm{VZ}(\xi)$ is a zero-duration virtual gate $Z_{\xi}$ for subsequent gates $S_{\gamma}S_\delta$ with $\xi=\gamma-\delta$ \cite{zgates}.}
\label{wav}
\end{figure}

Physically the experiment is sequence of preparation in the state $|0\rangle$, two gates $S_\alpha$, $S_\beta$ for the preparation,
two gates $S_\phi$, $S_\theta$, and the the readout pulse for the measurement of the state $|0\rangle$ again, see Fig. \ref{gates}. There are $5$ pairs of 
angles $\alpha_j,\beta_j$ to be chosen independently of the $4$ pairs $\theta_k,\phi_k$. Then $N=S_\beta S_\alpha|0\rangle\langle 0|$
and $M=S^\dag_\phi S^\dag_\theta|0\rangle\langle 0|S_\theta S_\phi$. The actual pulse  waveform of a sample sequence of gates is depicted in Fig. \ref{wav}.

\section{Experiment}

\begin{figure}
\includegraphics[scale=.7]{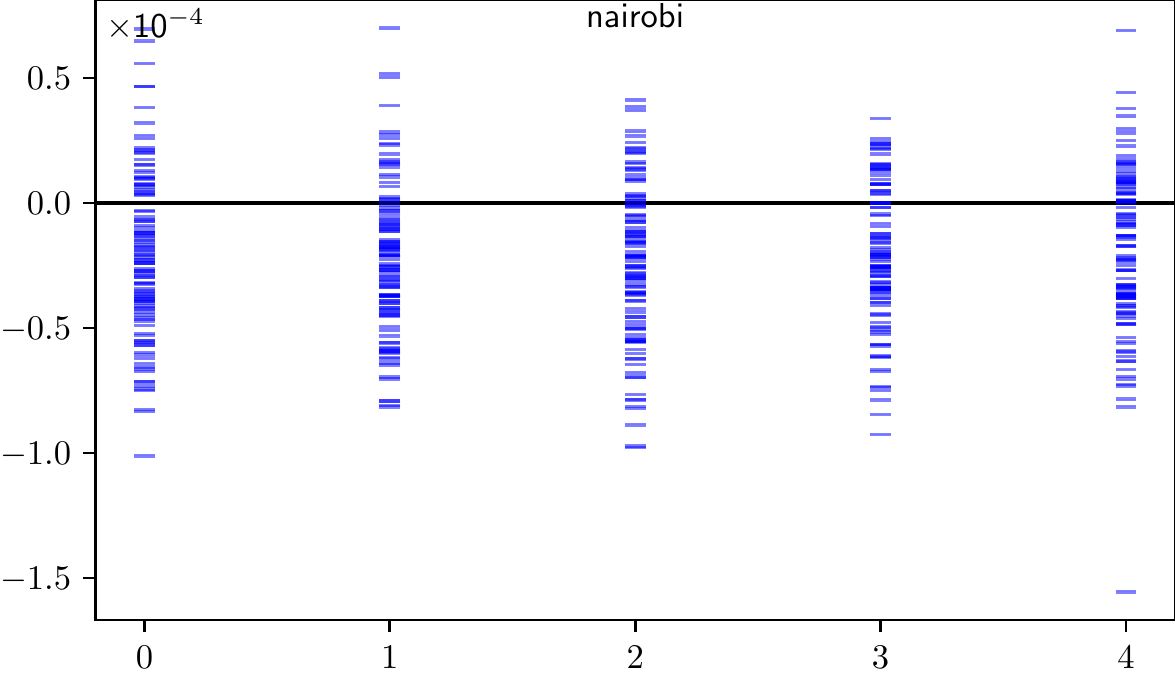}
\includegraphics[scale=.7]{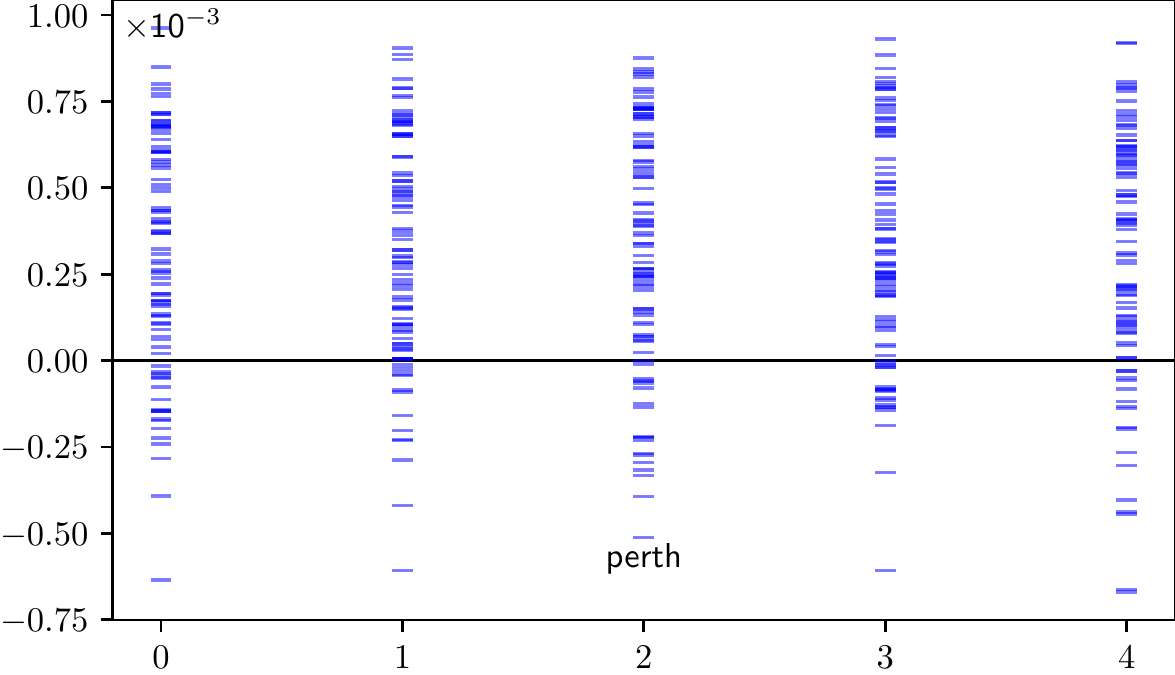}
\caption{Results of the test (II) the witness $W=\det p$, for the angles from Table \ref{tab2}, for nairobi and perth, for individual jobs. 
Two values for nairobi are beyond the picture boundaries, $(3,0.0023)$ and $(4,0.003)$}
\label{scat}
\end{figure}

In a perfect theory, we can predict a probability for every choice of $\alpha,\beta,\theta,\phi$. The experimental results can differ for a variety of
reasons. Firstly, the test is random and we have to estimate the error due to finite statistics.
For $T$ times the experiment is repeated, the variance of $W$ can be estimated as
\be
T\langle W^2\rangle\simeq \sum_{kj}p_{kj}(1-p_{kj})(\mathrm{Adj}\: p)_{jk}^2,\label{err}
\ee
where Adj is the adjoint matrix (matrix of minors of $p$, with crossed out a given row and column, and then transposed). Note that the identity 
$p^{-1}\det p=\mathrm{Adj}p$ makes no sense here as $W=\det p=0$ in the limit $T\to\infty$.
Secondly, the implementation of gates may be not faithful. Our test is capable to take them into account as long as the leakage to external states (e.g. $|2\rangle$)
is incoherent and does not depend on the parameters $\alpha,\beta,\theta,\phi$. Lastly, we have to assume that the pulse does not depend on the previous ones.
In other words, we can only test the combination of assumptions, dimension of the space and independence of operations.
We have calculated $W$ in two ways, (i) determining $p$ for each job and then 
finding $W$ (see the values for each job in Fig. \ref{scat}) and finally averaging $W$, (ii) averaging first $p$ from all jobs and then finding $W$.

\begin{figure}
\includegraphics[scale=.5]{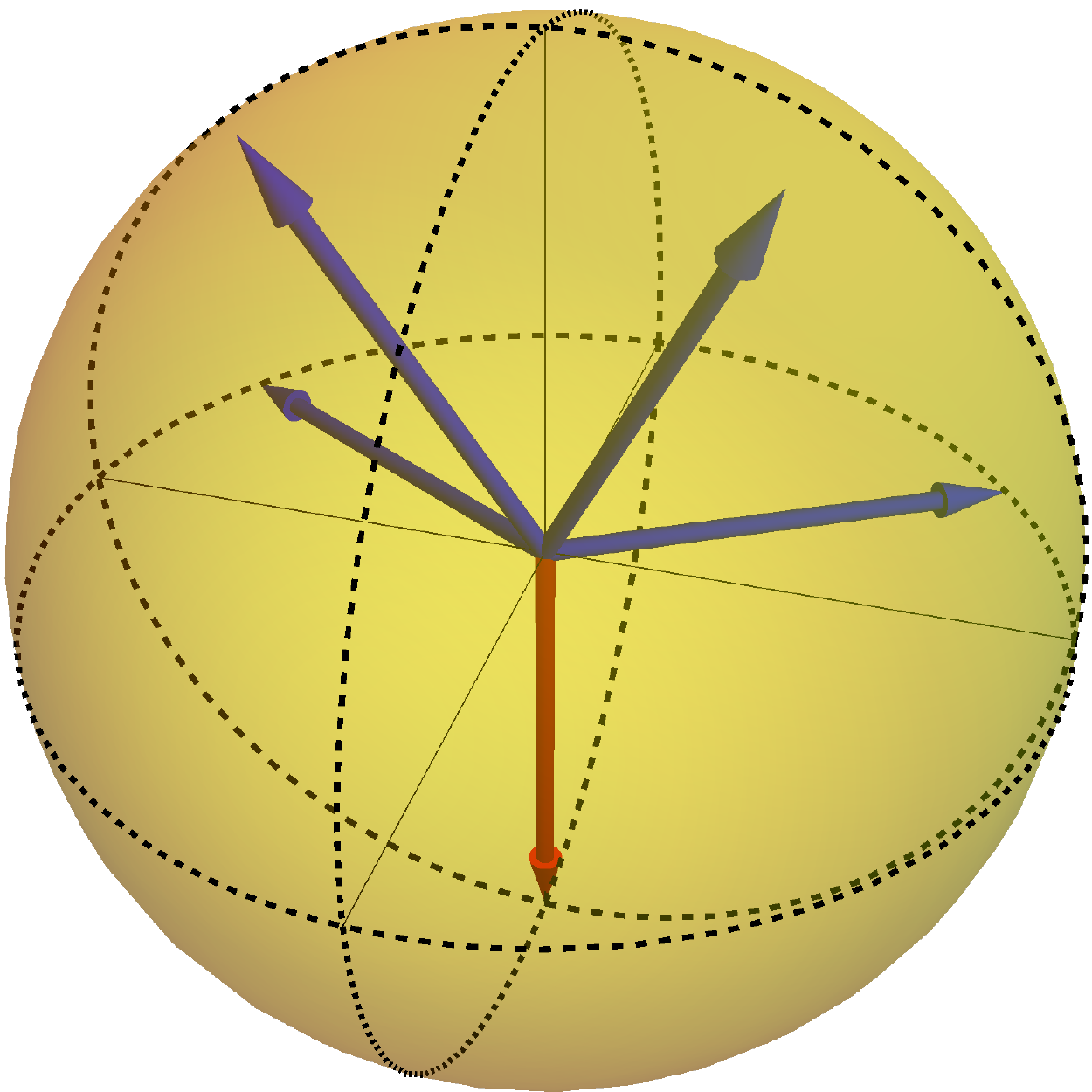}
\includegraphics[scale=.5]{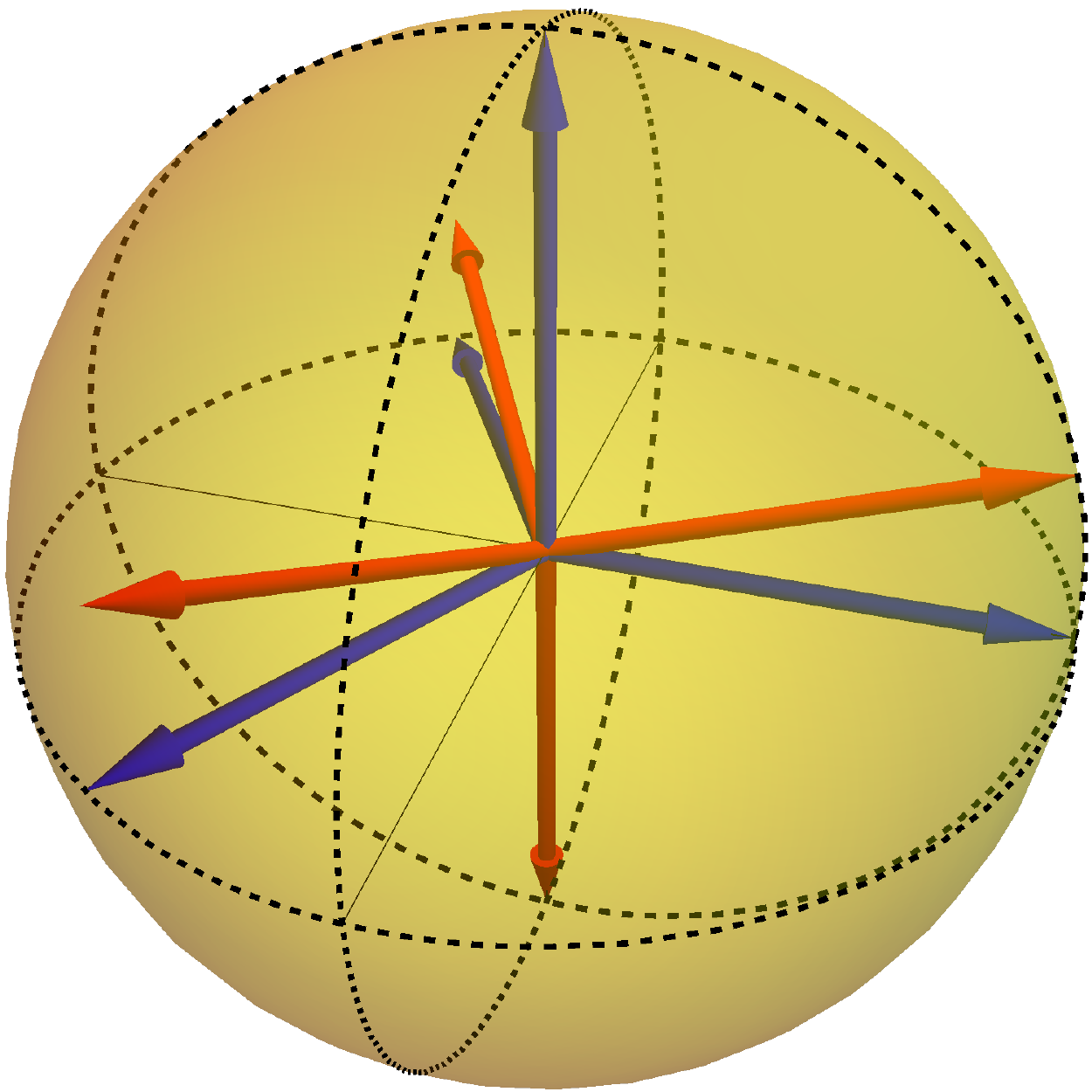}
\caption{The Bloch vectors for the preparations (red) and measurements (blue) corresponding to the angles from Table \ref{tab1}, top ' and bottom ''.
For the case ', the four measurement direction are identical to four preparations.}
\label{bloch1}
\end{figure}

\begin{figure}
\includegraphics[scale=.8]{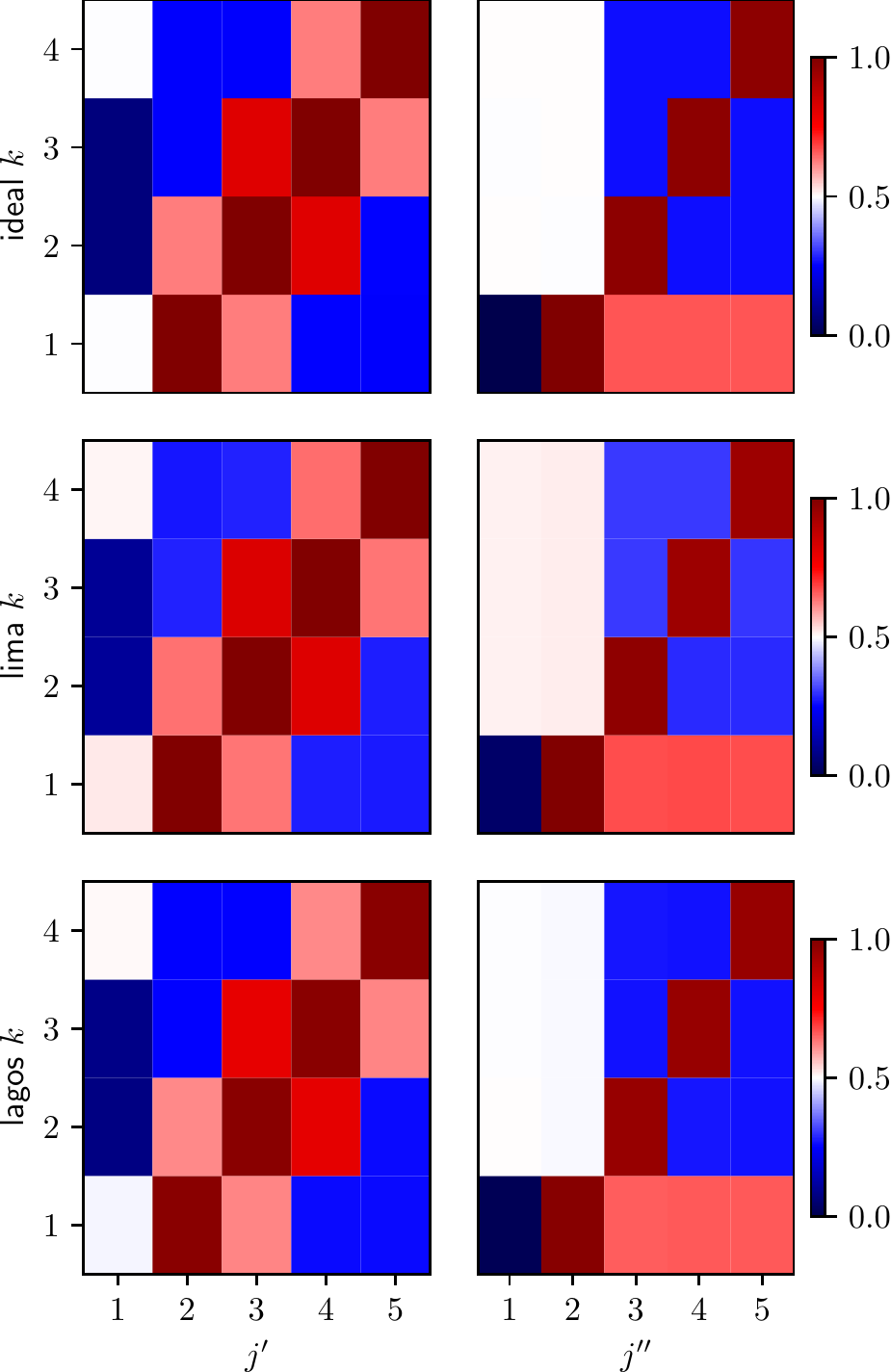}
\caption{Results of the test (I) with probabilities $p_{kj}$ for the angles from Table \ref{tab1}, for lima and lagos, compared to the ideal expectation.
Lagos: 60 jobs, 32000 shots, 15 repetitions. Lima: 521'/194'' jobs, 20000 shots, 5 repetitions}
\label{res1}
\end{figure}

\begin{figure}
\includegraphics[scale=.7]{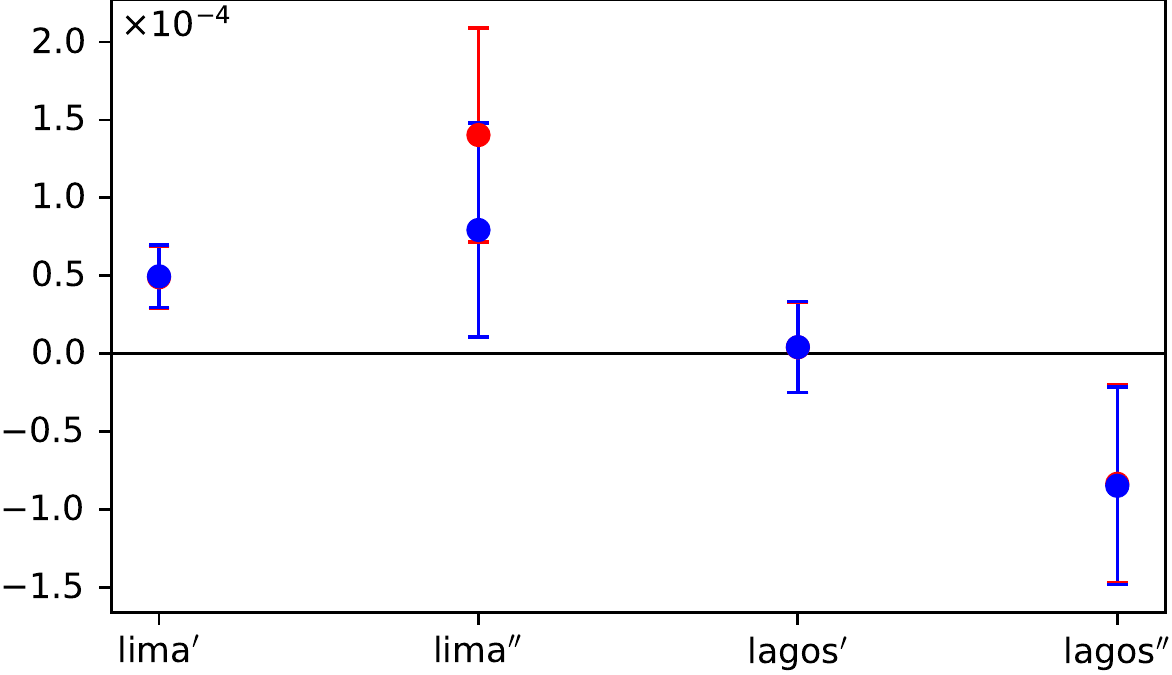}
\caption{Results of the test (I) the witness $W=\det p$, for the angles from Table \ref{tab1}, for lima and lagos, with the error given by (\ref{err}).
Red -- $W$ for $p$ from each job and then averaged, blue -- $p$ averaged from all jobs to give $W$.}
\label{dev1}
\end{figure}

There is no a priori best selection of preparations and measurements but they should not lie on a single Bloch circle.  We decided to make two kinds of tests:
(I) two special configurations corresponding to either the same Bloch vectors for preparation and measurements or maximal $\langle W^2\rangle$ for a given $R$;
(II) a family of configurations with one preparation vector at one of the 5 directions on the Bloch circle.
In both cases the corresponding Bloch vectors are derived explicitly in Appendix \ref{apa}.
The sets of angles in the case (I) are given in the Table \ref{tab1}, and the corresponding Bloch vectors are visualized in Fig. \ref{bloch1}.
We have run the test on lima and lagos, qubit 0. The probability matrix, compared to the ideal expectation is depicted in Fig. \ref{res1}.
The deviation from zero and the statistical error is given in Fig. \ref{dev1}. The number of  $T=\#\mathrm{jobs}\cdot \#\mathrm{shots}\cdot\#\mathrm{repetitions}$.
Technically, one sends a list of jobs to execute, each job contains up to $300$ circuits, to be distributed between experiments repeated the same number of times.
Each circuit is run the number of shots. The readout counts for each circuit is the value returned after the job execution is accomplished.

\begin{figure}
\includegraphics[scale=.5]{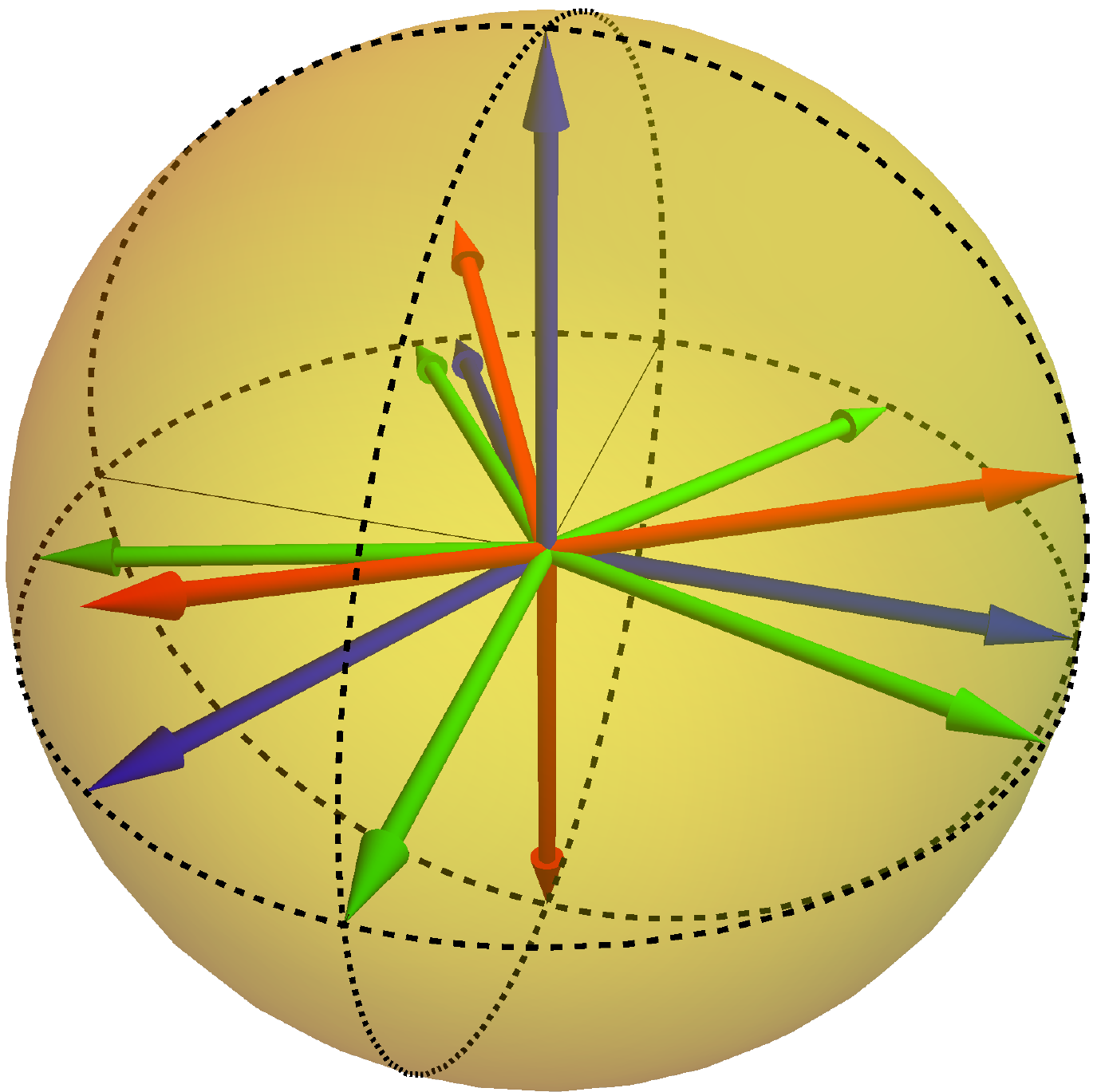}
\caption{The Bloch vectors for the parametric case (II) with fixed preparations (red) and measurements (blue) corresponding to the angles from Table \ref{tab1},
and a parametric preparation (green).}
\label{bloch2}
\end{figure}

\begin{figure}
\includegraphics[scale=.6]{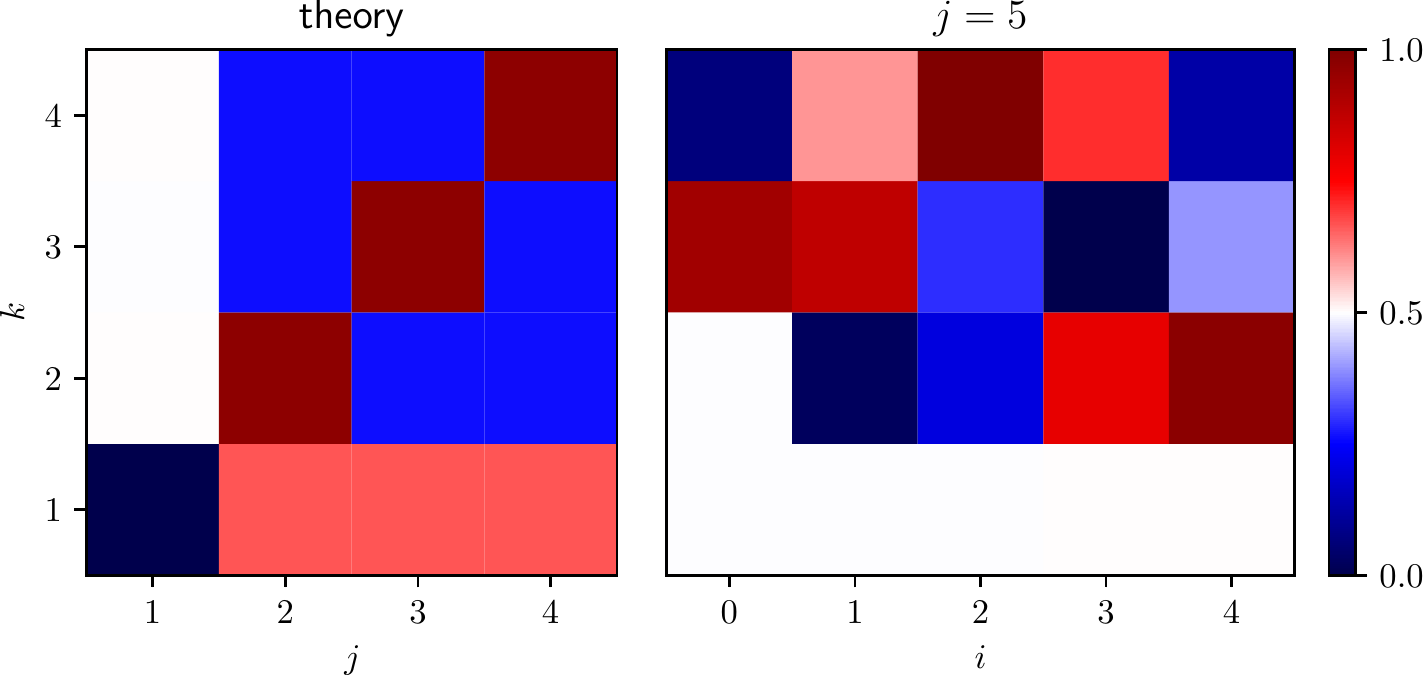}
\includegraphics[scale=.6]{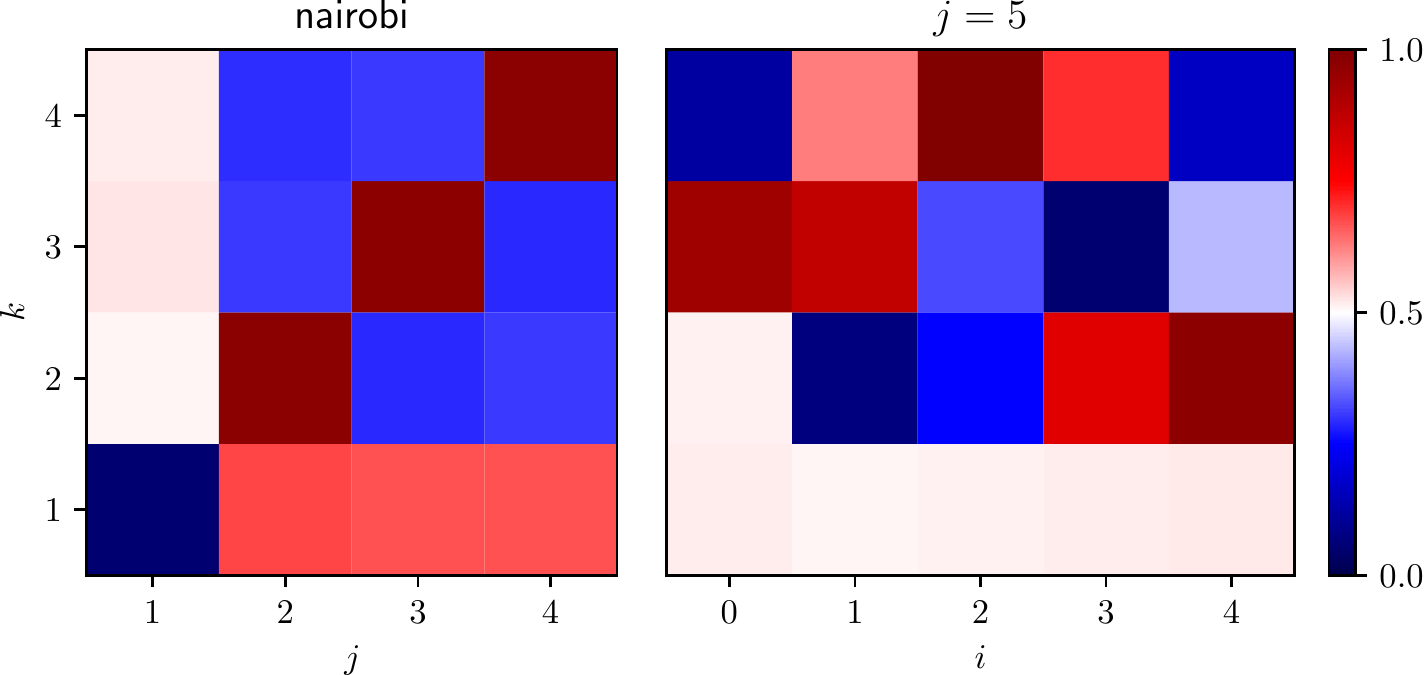}
\includegraphics[scale=.6]{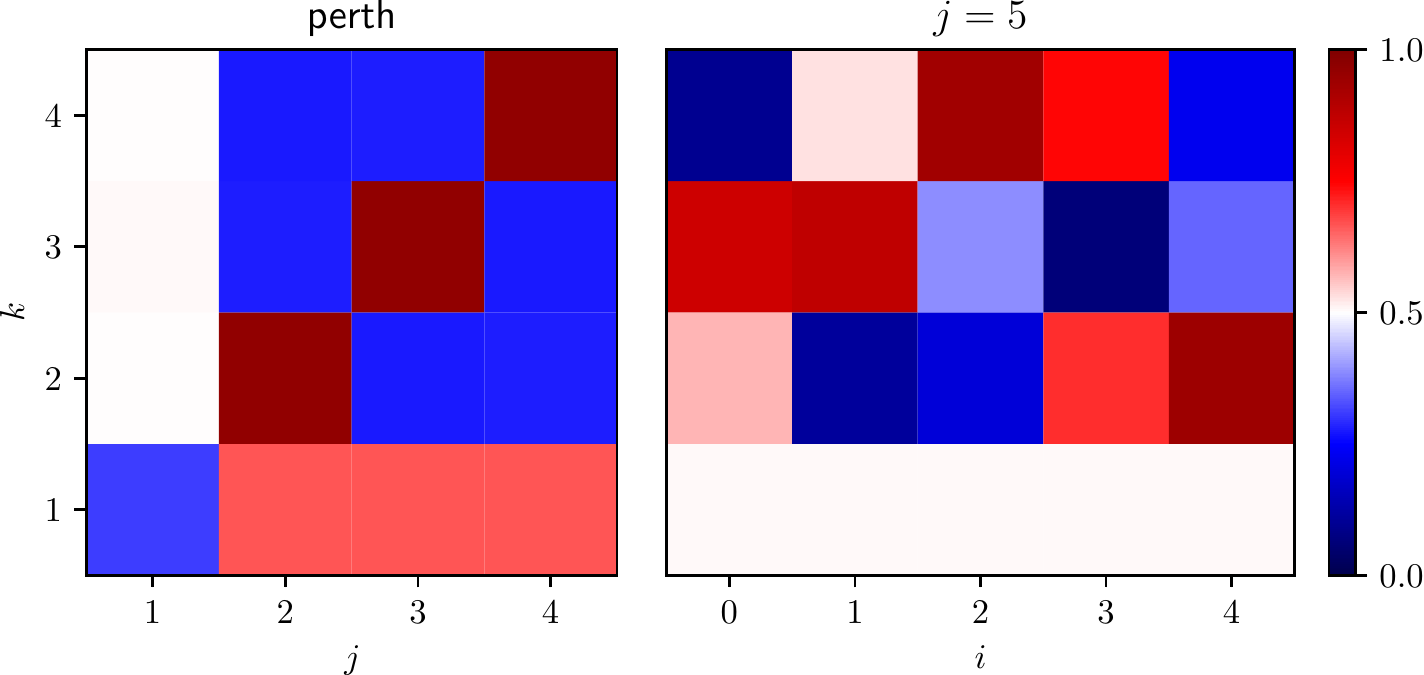}
\caption{Results of the test (II) with probabilities $p_{kj}$ for the angles from Table \ref{tab2}, for nairobi and perth, compared to the theory expectation.
Nairobi/perth: 115/93 jobs, both 100000 shots and 8 repetitions}
\label{res2}
\end{figure}

\begin{figure}
\includegraphics[scale=.7]{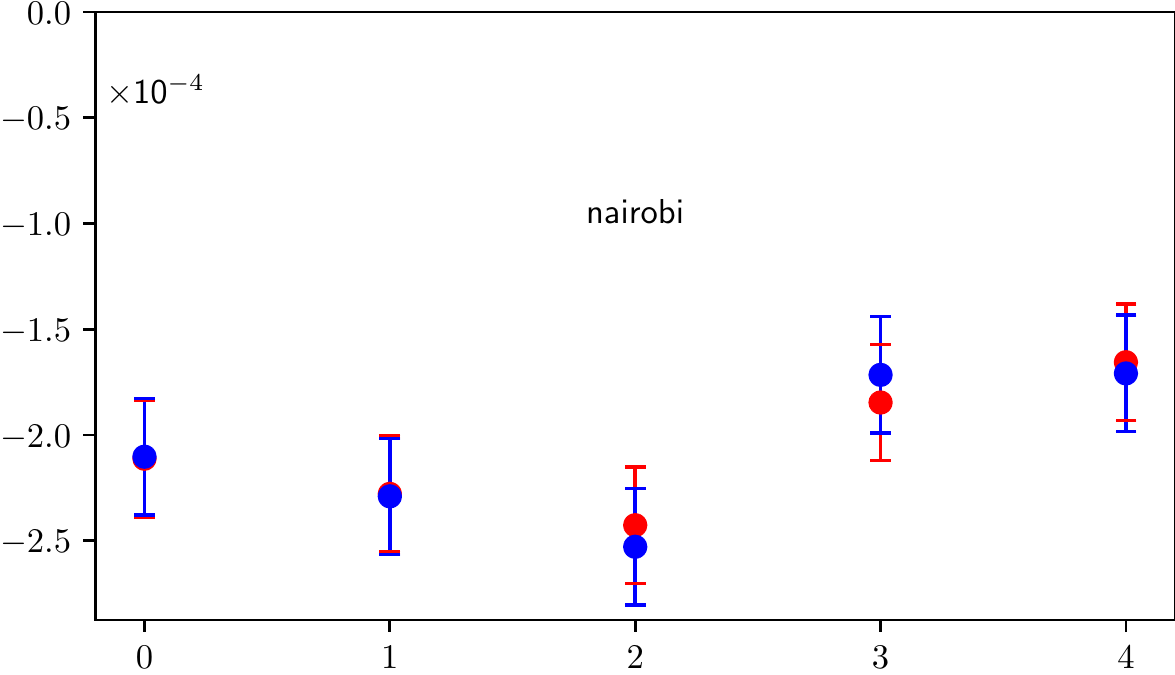}
\includegraphics[scale=.7]{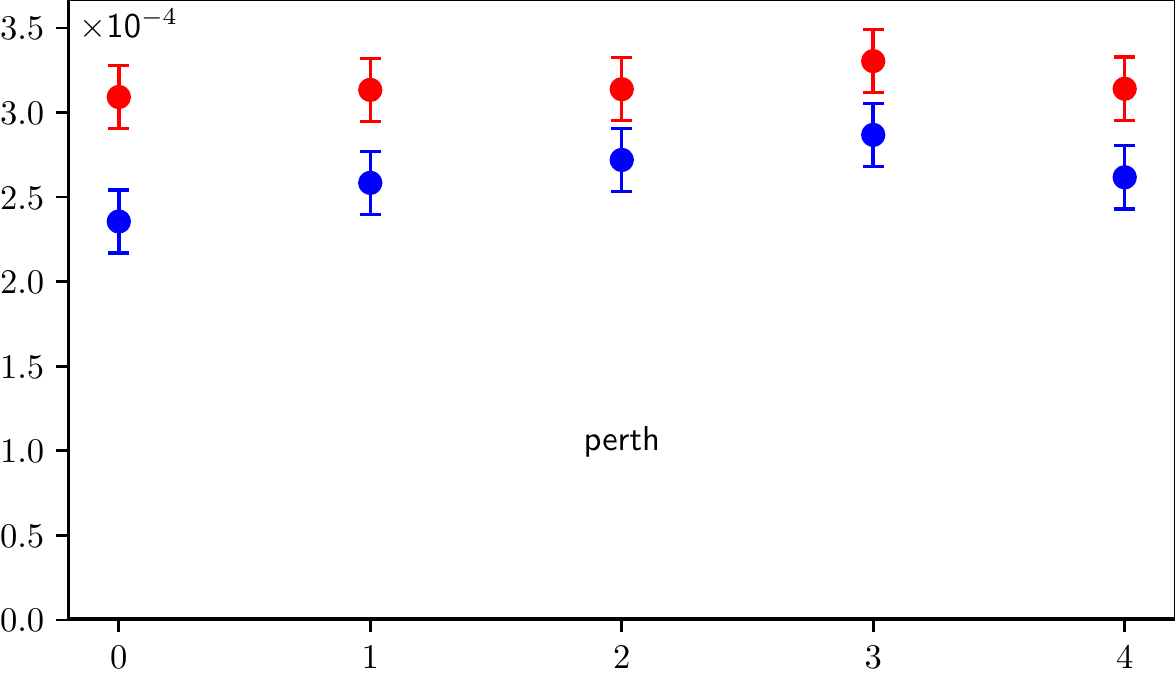}
\caption{Results of the test (II) the witness $W=\det p$, for the angles from Table \ref{tab2}, for nairobi and perth, with the error given by (\ref{err}).
Red -- $W$ for $p$ from each job and then averaged, blue -- $p$ averaged from all jobs to give $W$.}
\label{dev2}
\end{figure}

The sets of angles in the case (II) are prepared differently. Four preparations and measurements are fixed while the last preparation is parameter-dependent.
The fixed angles are specified in Table \ref{tab2}. The last preparation angles are $\alpha_5=2\pi i/5=\beta_5-\pi/2$ for $i=0..4$. 
The corresponding Bloch vectors are depicted in 
Fig. \ref{bloch2}. We have run the test on nairobi and perth, qubit 0. The probability matrix, compared to the ideal expectation is depicted in Fig. \ref{res2}.
The deviation from zero and the statistical error is given in Fig. \ref{dev2}. 
The deviation from the expected $0$ is more than $5$ standard deviations. The data and scripts are available at the public repository \cite{zen}.

\begin{table}
\begin{tabular}{*{10}{c}}
\toprule
$j$&1&2'&3'&4'&5'&2''&3''&4''&5''\\
\midrule
$\alpha$&0&$2\pi/3$&$2\pi/3$&$4\pi/3$&$4\pi/3$&$0$&$\eta-\pi$&$\eta+5\pi/3$&$\eta+\pi/3$\\
$\beta$&0&$\pi/6$&$-\pi/6$&$\pi/6$&$-\pi/6$&$\pi$&$0$&$2\pi/3$&$-2\pi/3$\\
\bottomrule
\end{tabular}

\begin{tabular}{*{9}{c}}
\toprule
$k$&1'&2'&3'&4'&1''&2''&3''&4''\\
\midrule
$\theta$&$5\pi/3$&$5\pi/3$&$\pi/3$&$\pi/3$&$\pi$&$\pi/2$&$7\pi/6$&$-\pi/6$\\
$\phi$&$7\pi/6$&$5\pi/6$&$7\pi/6$&$5\pi/6$&$0$&$\pi$&$5\pi/3$&$\pi/3$\\
\bottomrule
\end{tabular}
\caption{The angles for the special two special cases, ' and '', with $\eta=\acos(1/3)$ and 1=1'=1''.}
\label{tab1}
\end{table}

\begin{table}
\begin{tabular}{*{5}{c}}
\toprule
$j$&1&2&3&4\\
\midrule
$\alpha$&0&$\eta-\pi$&$\eta+5\pi/3$&$\eta+\pi/3$\\
$\beta$&0&$0$&$2\pi/3$&$-2\pi/3$\\
\bottomrule
\end{tabular}
\begin{tabular}{*{9}{c}}
\toprule
$k$&1&2&3&4\\
\midrule
$\theta$&$\pi$&$\pi/2$&$7\pi/6$&$-\pi/6$\\
$\phi$&$0$&$\pi$&$5\pi/3$&$\pi/3$\\
\bottomrule
\end{tabular}
\caption{The angles for the parametric case (II) for preparations and measurements 1..4}
\label{tab2}
\end{table}

\subsection{Nonidealities}

There are several factors that can affect the correctness of the experiment. (A) The daily calibration. The drive frequency and the gate
waveforms are corrected so different jobs can rely on different realizations of gates. There first order effect of calibrations is cancelled out. 
Nevertheless, we made more detailed estimates on second order effects in Appendix \ref{apb}.
Only large, unexpected failures could be a problem. (B) Amplitude-dependent leakage and distortion of the waveform. 
The leakage to higher states, e.g. $|2\rangle$ is small, of the order $10^{-4}$ and incoherent \cite{leak,praca}, see details in Appendix \ref{apc}. It is possible that 
distortion of amplitude to the waveform depends on the rotation angle (phase) but we expect this
effect to be very small, $10^{-3}$, based on the deviations observed in our previous work, and so the net effect is $10^{-7}$.
(C) Memory of the waveform between successive gates. Highly unlikely, a residual voltage amplitude can persist up to the next gate. In principle it can be mitigated
by delay-separated gates if the effect fades out with time. (D) Other qubits. They are usually detuned but some crosstalk may remain. 
As in the case of leakage, we expect the crosstalk to be incoherent and so irrelevant for the witness. 
As a sanity check we have run simulations, using the noise models from nairobi and perth, and no significant deviation have been found, see Appendix \ref{apd}.

\section{Discussion}
A test of linear independence of quantum operations reveals subtle deviations, invisible in more crude tests. 
Further tests are necessary to identify the origin of the deviations, to exclude e.g. exotic many world/copies theories \cite{plaga,adp}.  We suggest: (i) an extreme statistics collected in a relatively short time to
avoid corrections due to calibrations, (ii) a time separation between gates to exclude potential overlap of the effects, (iii) a scan through a large set of 
Bloch vectors to maximize the potential deviation, (iv) run the test on a single-qubit devices to avoid cross-talks. It is also possible to
develop more sophisticated tests, with different assumptions, or involving different qubits. In any case, a precise diagnostics of qubits 
must become a standard in quantum technologies.

\appendix

\begin{figure}
\includegraphics[scale=.6]{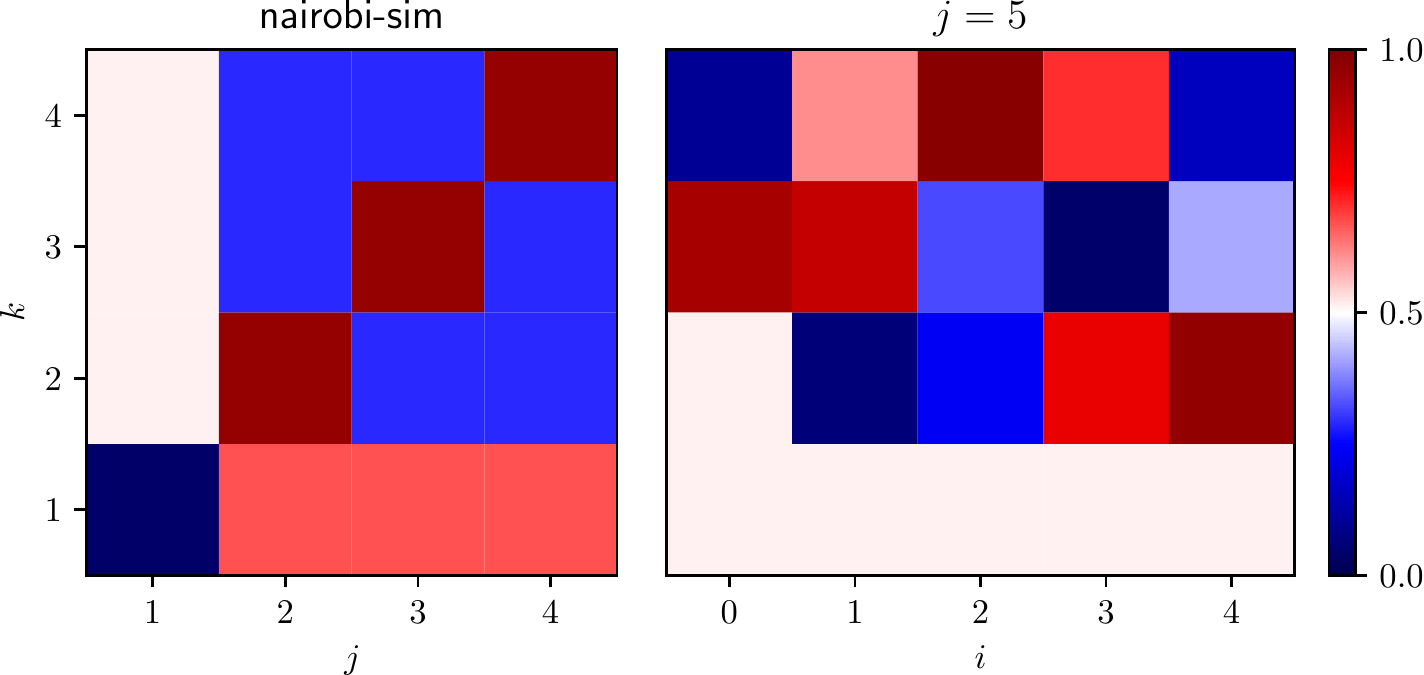}
\includegraphics[scale=.6]{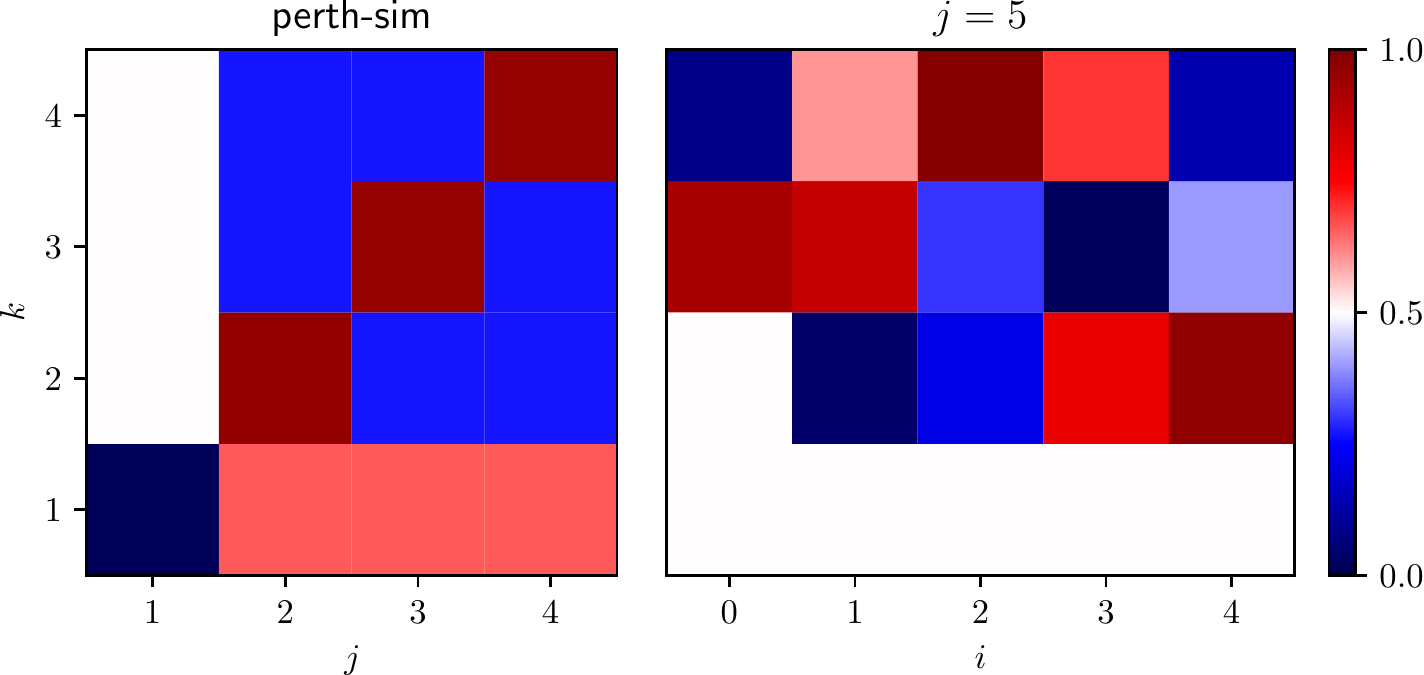}
\caption{Results of the simulations of the test (II) with probabilities $p_{kj}$ for the angles from Table \ref{tab2}, for nairobi and perth.}
\label{res2-s}
\end{figure}

\begin{figure}
\includegraphics[scale=.7]{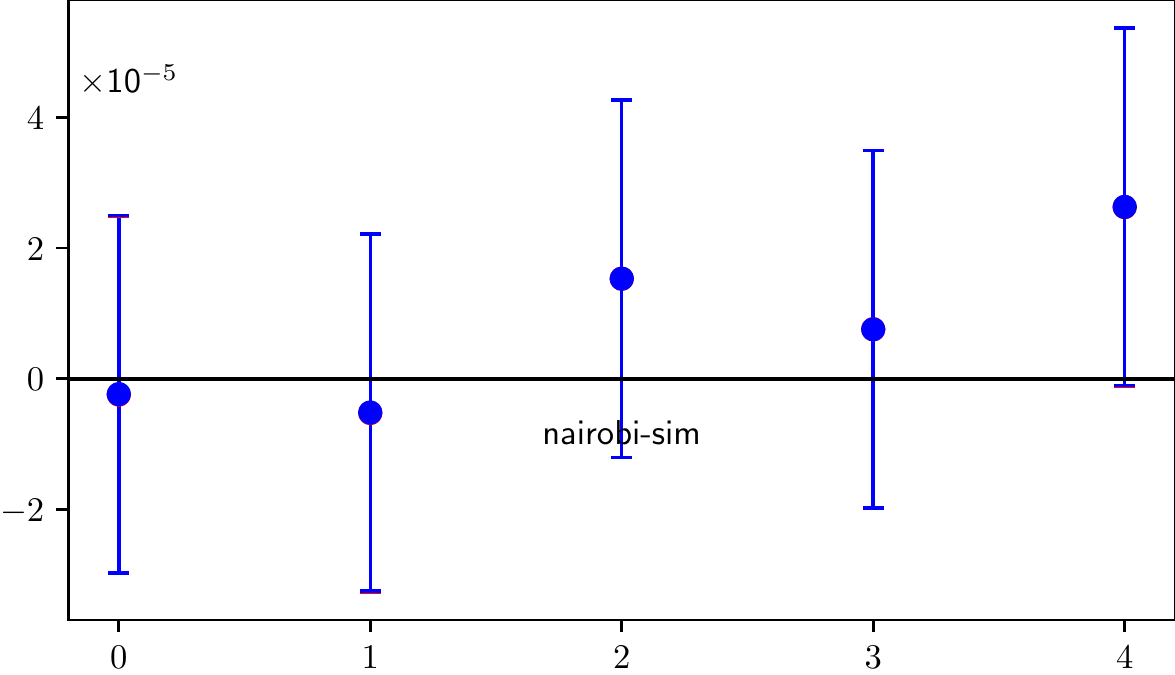}
\includegraphics[scale=.7]{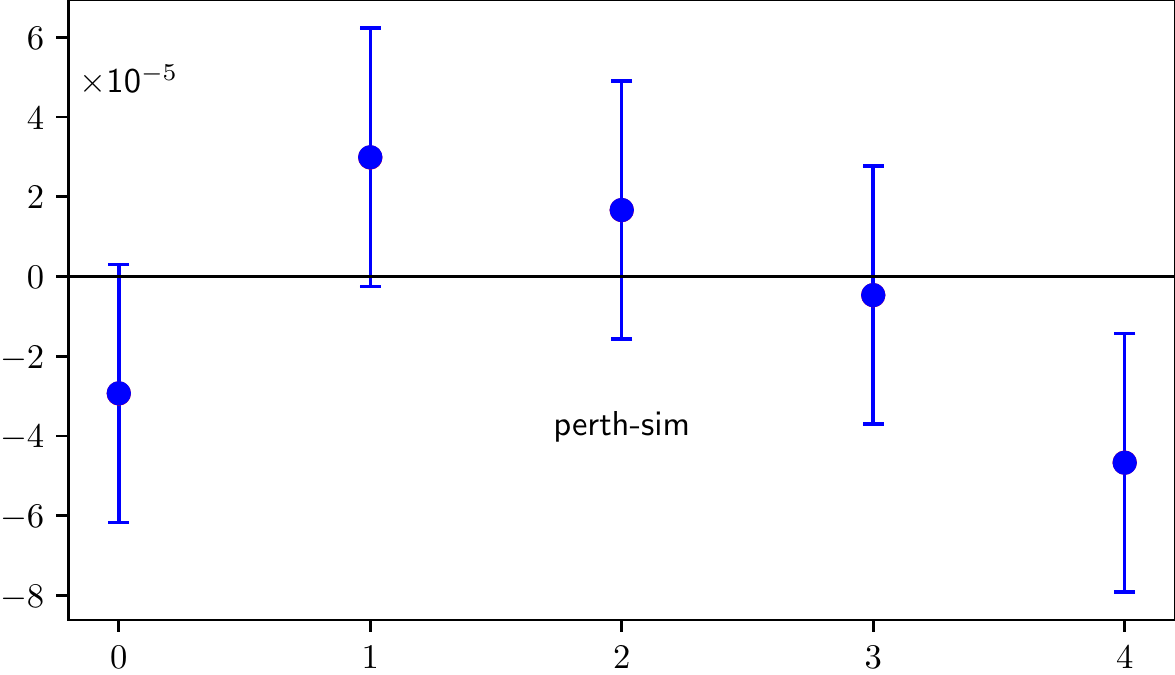}
\caption{Results of the simulations of the test (II) the witness $W=\det p$, for the angles from Table \ref{tab2}, for nairobi and perth noise models.
Note that the two ways of calculation of $W$ almost coincide (the blue one covers the red one), which is consistent with our explanation
of averaged out first order difference in Appendix \ref{apb}.}
\label{dev2-s}
\end{figure}

\begin{figure}
\includegraphics[scale=.7]{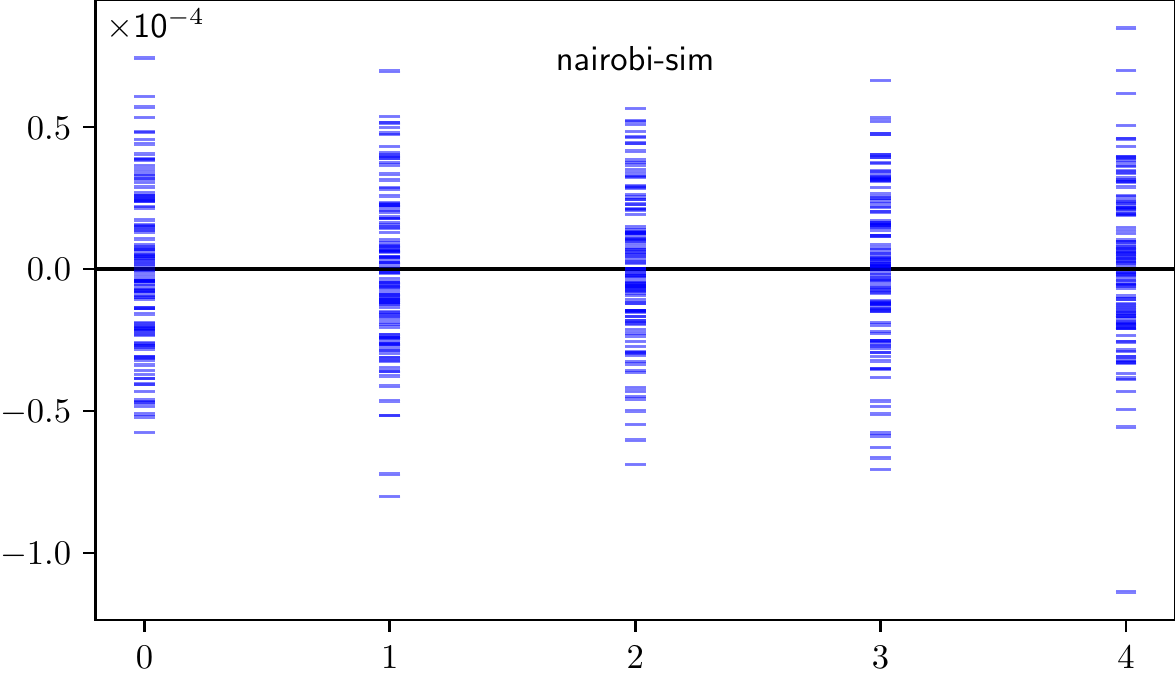}
\includegraphics[scale=.7]{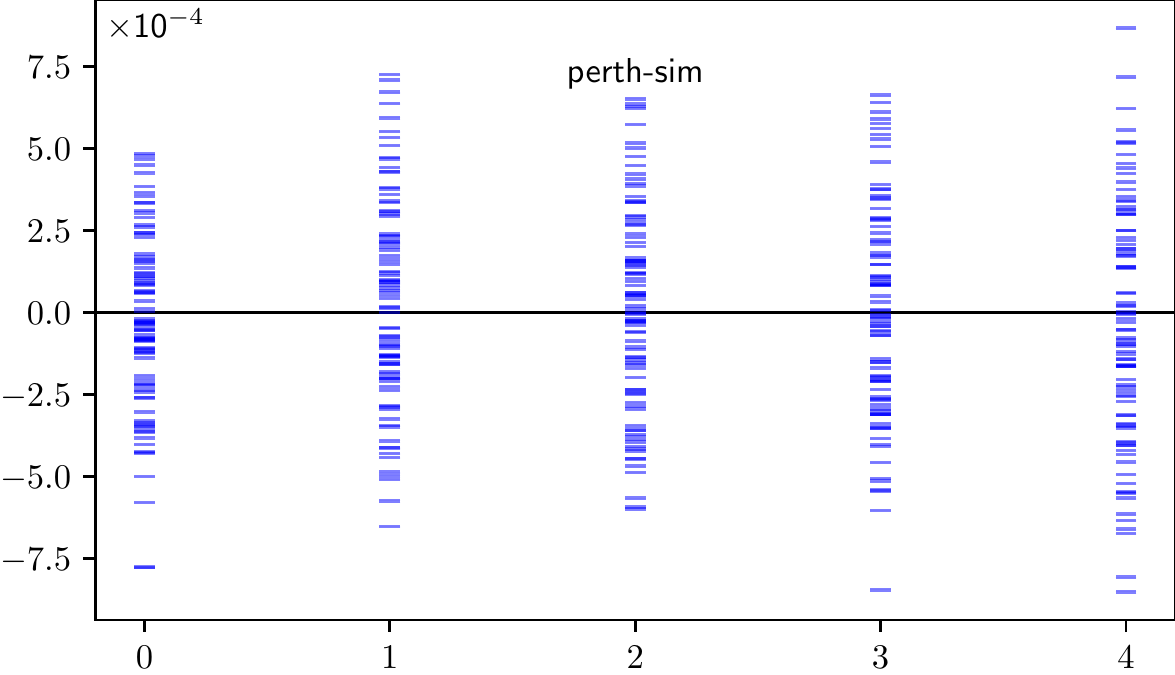}
\caption{Results of the simulations of the test (II) the witness $W=\det p$, for the angles from Table \ref{tab2}, for nairobi and perth noise models, for individual jobs}
\label{scat-s}
\end{figure}

\section{Bloch sphere representations}
\label{apa}

Using vectors $\boldsymbol n$ to represent the state $N=|\boldsymbol n\rangle\langle \boldsymbol n|=(\hat{1}+\boldsymbol n\cdot\boldsymbol \sigma)/2$,
we have $S_\alpha N S^\dag_\alpha=N_\alpha$
and $S_\theta^\dag M S_\theta=M_\theta$ with
\begin{align}
&\boldsymbol n_\alpha=\begin{pmatrix}
\cos^2\alpha&-\cos\alpha\sin\alpha&-\sin\alpha\\
-\cos\alpha\sin\alpha&\sin^2\alpha&-\cos\alpha\\
\sin\alpha&\cos\alpha&0\end{pmatrix}\boldsymbol n,\nonumber\\
&\boldsymbol m_\theta=\begin{pmatrix}
\cos^2\theta&-\cos\theta\sin\theta&\sin\theta\\
-\cos\theta\sin\theta&\sin^2\theta&\cos\theta\\
-\sin\theta&-\cos\theta&0\end{pmatrix}\boldsymbol n,
\end{align}
For $\boldsymbol n=\boldsymbol m=(0,0,1)$ and $m_0=1$, we have
$N_{\alpha\beta}=S_\beta S_\alpha N S^\dag_\alpha S^\dag_\beta$ with
\be
\boldsymbol n'_{\alpha\beta}=(\sin(\beta-\alpha)\cos\beta,\sin(\alpha-\beta)\sin\beta,-\cos(\beta-\alpha))
\ee
while $M_{\theta\phi}=S^\dag_\phi S^\dag_\theta MS_\theta S_\phi$ wirh
\be
\boldsymbol M_{\theta\phi}=(\sin(\theta-\phi)\cos\phi,\sin(\phi-\theta)\sin\phi,-\cos(\theta-\phi)).
\ee
Then the probability matrix elements read
\begin{equation}
p_{kj}=\mathrm{Tr}M_kN_j=(1+\boldsymbol n\cdot\boldsymbol m)/2
\end{equation}
while $p_{5j}=1$.

In this way we can represent the choices used in our experiment. In the first choice, preparations 
$\boldsymbol n'_1=(0,0,-1)$, $\boldsymbol n'_2=(-\sqrt{3}/2,1/2,0)$, $\boldsymbol n'_3=(-\sqrt{3}/4,-1/4,\sqrt{3}/2)$,
$\boldsymbol n'_4=(\sqrt{3}/4,-1/4,\sqrt{3}/2)$, $\boldsymbol n'_5=(\sqrt{3}/2,1/2,0)$ and measurements $\boldsymbol m'_k=\boldsymbol n'_{k-1}$.
In the second choice,
$\boldsymbol n''_1=-\boldsymbol n''_2=(0,0,-1)$, 
$\boldsymbol n''_3=(2\sqrt{2},0,1/3)$, $\boldsymbol n''_{4,5}=(-\sqrt{2}/3,\mp\sqrt{2/3},1/3)$
and measurements $\boldsymbol m''_1=(0,0,1)$, $\boldsymbol m''_2=(1,0,0)$, $\boldsymbol m''_{3,4}=(-1/2,\mp\sqrt{3}/2,0)$.

For the parametric test we have
$\boldsymbol n_1=(0,0,1)$,
$\boldsymbol n_2=(2\sqrt{2},0,1/3)$, $\boldsymbol n_{3,4}=(-\sqrt{2}/3,\mp\sqrt{2/3},1/3)$ while
$\boldsymbol n_5^i=(-\sin(2\pi i/5),-\cos(2\pi i/5),0)$.

\section{Bounds on daily calibrations}
\label{apb}

Suppose that the calibration from job to job can alter the matrix of probabilities. Assuming that each job $n=1..N$ satisfies $W^{(n)}=0$ for probabilities 
$p^{(n)}$, we ask
if $W$ for $p=\sum_n p^{(n)}/N$ can be nonzero. Suppose $\delta p^{(n)}=p^{(n)}-p^{(0)}$ is small for some reference matrix
$p^{(0)}$  and $|\delta p_{kj}^{(n)}|\leq \epsilon$ for all ${kj}$ and some small bound $\epsilon$.
Then, in the first order of $\delta p$ we have still
$W\simeq 0$ from expanding determinant in linear combinations of single columns $p^{(n)}$ and the rest of columns kept equal $p^{(0)}$.
The nonvanishing contribution is of the second order, when replacing either of two columns by $\delta p^{(n)}$. Their length is $\leq 2\epsilon$.
The last row contains $0$ for the replaced columns and $1$ for the rest. Subtracting $1/2$ of that row from the other rows. The 
moduli of remaining elements are $\leq 1/2$ for the length of the remaining $3$ columns is $\leq \sqrt{2}$.
 From Hadamard inequality $|\det A|\leq \prod_j |A_j|$ with $|A_j$ being the length of the vector
(column) $A_j$ of the matrix $A$, we have the upper bound $|W|\leq 80\sqrt{2}\epsilon^2$ as we have $10$ choices of $2$ columns out of $5$.

\section{Corrections from higher states}
\label{apc}

The generic Hamiltonian, in the basis states $|n\rangle$, $n=0,1,2,...$ ($\hbar=1$) reads

\be
H=\sum_n \omega_n|n\rangle\langle n|+2\cos(\omega t-\theta)\hat{V}(t)
\ee
with energy  $\omega_n$ eigenstates levels  and the external drive $V$ at frequency $\omega$ and phase shift $\theta$  (the second term). In principle free parameters $\omega,\theta$ and $\hat{V}(t)$ can model a completely arbitrary evolution.
We can estimate deviations by perturbative analysis, setting $\omega_0=0$, $\omega_1=\omega$ (resonance), 
$\omega_2=2\omega+\omega'$ (anharmonicity, i.e. $\omega'\ll\omega$, in IBM about $300$Mhz compared to drive frequency $\sim 5$GHz). 
The state $|2\rangle$ should give the most significant potential contribution.
We can incorporate rotation and phase into the definition of states, $|n\rangle\to |n'\rangle=e^{-in(\theta+\omega t)}|n\rangle$ so that
\begin{widetext}
\be
H'=
\mb
2\cos(\omega t-\theta)V_{00}&(1+e^{-2i(\theta+\omega t)}) V_{01}&(e^{-i(\theta+\omega t)}+e^{-3i(\theta+\omega t)}) V_{02}\\
(1+e^{2i(\omega t+\theta)})V_{10}&2\cos(\omega t+\theta)V_{11}&
(1+e^{-2i(\theta+\omega t)}) V_{12}\\
(e^{-i(\theta+\omega t)}+e^{3i(\omega t+\theta)}) V_{20}&
(1+e^{2i(\omega t+\theta)}) V_{21}&
2\cos(\omega t+\theta)V_{22}+\omega'
\me
\ee
Extracting  the Rotating Wave Approximation (RWA) part from $H'=H_{RWA}+\Delta H$,
\be
H_{RWA}=
\mb
0&V_{01}&0\\
V_{10}&0&
 V_{12}\\
0&
V_{21}&
\omega'
\me,
\ee
the correction reads
\be
\Delta H=\mb
2\cos(\omega t+\theta)V_{00}&e^{-2i(\theta+\omega t)} V_{01}&(e^{-i(\theta+\omega t)}+e^{-3i(\theta+\omega t)}) V_{02}\\
e^{2i(\omega t+\theta)}V_{10}&2\cos(\omega t+\theta)V_{11}&
e^{-2i(\theta+\omega t)} V_{12}\\
(e^{-i(\theta+\omega t)}+e^{3i(\omega t+\theta)}) V_{20}&
e^{2i(\omega t+\theta)} V_{21}&
2\cos(\omega t+\theta)V_{22}
\me
\ee
\end{widetext}
Evolution due to RWA has the form
\be
U(t)=\mathcal T\exp\int_{-\infty}^t  H_{RWA}(t')dt'/i,
\ee
where $\mathcal T$ means chronological product in Taylor expansion.
Then the 1st order correction to $U$ reads
\be
\Delta U=U(+\infty)\int dt U^\dag(t)\Delta H(t)U(t)/i
\ee
where the full rotation is $U(+\infty)$.
All $\theta$-dependent terms in $\Delta H$, contain also $e^{i\omega t}$, which exponentially damps slow-varying expressions.
The 2nd order correction reads
\begin{align}
&\Delta^2 U=-U(+\infty)\times\nonumber\\
&\int dt U^\dag(t)\Delta H(t)U(t)\int^t dt'U^\dag(t')\Delta H(t')U(t').
\end{align}
Most of components get damped exponentially, too, except when $\Delta H(t)$ contains $e^{ik\omega t}$
and $\Delta H(t')$ contains $e^{-ik\omega t}$, $k=1,2,3$, so $k\theta$ cancels.
The nonnegligible part of $\Delta^2 U$ is therefore independent of $\theta$ giving slowly Bloch-Siegert shift \cite{blochs}.
Stroboscopic corrections to RWA \cite{rwa} can be neglected due to a very short sampling time, $dt=0.222$ns,

\section{Simulations}
\label{apd}

As a cross-check of our test we have run the identical programs on IBM simulator of a quantum computer with the noise model
taken from the real devices perth and nairobi. However, in contrast to real devices, the results are in agreement with the theory as shown in Figs.
\ref{res2-s}, \ref{dev2-s}, \ref{scat-s}.

\end{document}